\documentclass[11pt]{article}
\usepackage{graphicx,epsfig}

\newcommand{\BABARPubYear}    {04}

\newcommand{\BABARProcNumber} {011}
\newcommand{\SLACPubNumber} {10515}
\newcommand{\LANLNumber} {0406082}

\def\PLB{{\em Phys. Lett.}  B}

\def\PRD{{\em Phys. Rev.} D}

\def\be{\begin{equation}}
\def\ee{\end{equation}}
\def\bea{\begin{eqnarray}}
\def\eea{\end{eqnarray}}

\RequirePackage{xspace}
\usepackage{relsize}
\def\babar{\mbox{\slshape B\kern-0.1em{\smaller A}\kern-0.1em
    B\kern-0.1em{\smaller A\kern-0.2em R}}}
\def\stwob{\ensuremath{\sin\! 2 \beta   }\xspace}
\def\ctwob{\ensuremath{\cos\! 2 \beta   }\xspace}
\def\Bbar    {\kern 0.18em\overline{\kern -0.18em B}{}\xspace}
\def\Bz      {\ensuremath{B^0}\xspace}
\def\Bzb     {\ensuremath{\Bbar^0}\xspace}
\def\Bu      {\ensuremath{B^+}\xspace}

\def\Bp      {\ensuremath{\Bu}\xspace}

\def\Bpm     {\ensuremath{B^\pm}\xspace}
\def\fz    {\ensuremath{f_0\!}\xspace}
\def\Kz    {\ensuremath{K^0}\xspace}
\def\KS    {\ensuremath{K^0_{\scriptscriptstyle S}}\xspace}
\def\KL    {\ensuremath{K^0_{\scriptscriptstyle L}}\xspace}
\def\Kp    {\ensuremath{K^+}\xspace}
\def\Km    {\ensuremath{K^-}\xspace}
\def\Kpm   {\ensuremath{K^\pm}\xspace}

\def\Kstarz  {\ensuremath{K^{*0}}\xspace}

\def\Kstar   {\ensuremath{K^*}\xspace}

\def\CP                {\ensuremath{C\!P}\xspace}
\def\to                 {\ensuremath{\rightarrow}\xspace}
\def\jpsi     {\ensuremath{{J\mskip -3mu/\mskip -2mu\psi\mskip 2mu}}\xspace}
\def\piz   {\ensuremath{\pi^0}\xspace}
\def\pip   {\ensuremath{\pi^+}\xspace}
\def\pim   {\ensuremath{\pi^-}\xspace}

\def\pimp  {\ensuremath{\pi^\mp}\xspace}

\long\def\inst#1{\par\nobreak\kern 4pt\nobreak
    {\it #1}\par\vskip 10pt plus 3pt minus 3pt}

\begin{document}
{\pagestyle{empty}

\begin{flushright}
SLAC-PUB-\SLACPubNumber \\
\babar-PROC-\BABARPubYear/\BABARProcNumber \\
hep-ex/\LANLNumber \\
June, 2004 \\
\end{flushright}

\par\vskip 4cm

\begin{center}
\Large \bf
Measurements Related to the CKM Angle $\beta/\phi_1$ from \babar
\end{center}
\bigskip

\begin{center}
\large 
M. Verderi \\
Laboratoire Leprince-Ringuet, Ecole polytechnique,\\
Route de Saclay, 91128 Palaiseau C\'edex, France \\
(for the \babar\ Collaboration)
\end{center}
\bigskip \bigskip

\begin{center}
\large \bf Abstract
\end{center}
We present measurements related to the CKM angle $\beta$ from \babar\ based
on 82 or $\sim$ 115 fb$^{-1}$ data samples collected at the PEP-II $e^+e^-$ asymmetric $B$ Factory collider.
A new method allowing to measure the {\em sign} of  \ctwob
with \Bz\to\jpsi``\Kstarz'' events is explained and related results reported.
Recent \CP measurements in the penguin dominated modes \Bz\to$\phi$\Kz, \Bz\to\Kp\Km\KS, \Bz\to\piz\KS and \Bz\to\fz$(980)$\KS are presented.

\vfill
\begin{center}
Contributed to the Proceedings of the
39$^{th}$ Rencontres De Moriond On Electroweak Interactions And Unified Theories, \\
03/21/2004---03/28/2004, La Thuile, Aosta Valley, Italy
\end{center}

\vspace{1.0cm}
\begin{center}
{\em Stanford Linear Accelerator Center, Stanford University, 
Stanford, CA 94309} \\ \vspace{0.1cm}\hrule\vspace{0.1cm}
Work supported in part by Department of Energy contract DE-AC03-76SF00515.
\end{center}

\section{Introduction}
The \CP violating parameter \stwob is measured with high precision, but this leaves a four--fold ambiguity on the angle $\beta$ itself. This can be reduced
to a two--fold ambiguity if measuring the sign of \ctwob. The measurement of this sign provides a direct test of the Standard Model (SM), since
$\ctwob > 0$ is expected in the SM. The \ctwob parameter can be measured with $\Bz\to\jpsi\Kstarz; \Kstarz\to\KS\piz$ events, but, precisely, with a sign ambiguity, coming
itself from a two--fold ambiguity in the determination of the strong phases involved in the decay. We present a new method, based on the analysis of the $K\pi$ phase variation
with mass, to break this strong phases ambiguity and the first ambiguity-free measurement of \ctwob, with  \Bz\to\jpsi``\Kstarz'' events.

The SM can be further challenged by \stwob measurements in penguin dominated channels, since new physics (NP) may enter the loop appearing in the diagram of these decays.
A large departure from the \stwob value measured with \Bz\to\jpsi\KS will indicate contribution of NP.
We present the recent \babar\ \CP measurements in \Bz\to$\phi$\Kz; \Bz\to\Kp\Km\KS with its companion channel \Bpm\to\Kpm\KS\KS, which a is new
\babar\ measurement; \Bz\to\piz\KS; and \Bz\to\fz$(980)$\KS, which is a first measurement.

\section{Measurement of the \ctwob sign with  \Bz\to\jpsi``\Kstarz'' events}
The \CP content of the decay  $\Bz\to\jpsi\Kstarz(892); \Kstarz(892)\to\KS\piz$ is both even an odd. The \ctwob parameter appears through \CP--even -- \CP--odd interferences in the
time and angular dependant distribution in the observables~\cite{steph}:
\begin{eqnarray}
\label{eqn:obs_cos}
\cos(\delta_\|-\delta_\perp)\cdot\ctwob,\;
\cos(\delta_\perp-\delta_0)\cdot\ctwob,
\end{eqnarray}
where $\delta_0, \delta_\|$ and $\delta_\perp$ are the strong phases of the decay amplitudes $A_0=|A_0|e^{i\delta_0}, A_\|=|A_\||e^{i\delta_\|}$, which are \CP--even,
and $A_\perp=|A_\perp|e^{i\delta_\perp}$, which is \CP--odd. These strong phases
are measured on a large sample of neutral and charged $B\to\jpsi K^*$ decays (tab.~\ref{tab:amp}), but up to the two-fold mathematical ambiguity
$(\delta_\|-\delta_0,\delta_\perp-\delta_0) \leftrightarrow (-(\delta_\|-\delta_0),\pi-(\delta_\perp-\delta_0))$.
\label{eqn:ambig}
\begin{table}[h]
\caption{\label{tab:amp}$B\to\jpsi K^*$ decay amplitude moduli (left) and strong phases (right) measured by an angular analysis on a sample
of $\Bz\to\jpsi(\Kp\pim)^{*0}, \Bp\to\jpsi(\KS\pip)^{*+}, \Bp\to\jpsi(\Kp\piz)^{*+}$, and related charged conjugate decays. The integrated luminosity is
$82$fb$^{-1}$. The yields corresponding to the three above channels are $2376\pm 51$, $670\pm 27$ and $791\pm 33$ respectively.
For the strong phases (right table), the values corresponding to the two ambiguous solutions (see text) are given.
Note that we observe a 7.6 $\sigma$ significant strong phase: $\delta_\| -\delta_\perp = 0.597\pm0.077\pm0.017$.}
\begin{minipage}{0.38\linewidth}
\[
\begin{array}{|c|c|}\hline
|A_0|^2 & 0.566\pm  0.012\pm  0.005\\
|A_\||^2 & 0.204\pm  0.015\pm  0.005\\
|A_\perp|^2 & 0.230\pm  0.015\pm  0.004\\\hline
\end{array}\hfill
\]
\end{minipage}
\begin{minipage}{0.60\linewidth}
\[
\begin{array}{|c|c|c|}\hline
 & {\rm Solution\ I} & {\rm Solution\ II} \\\hline
\delta_\| -\delta_0 & 2.729\pm  0.101\pm  0.052 & 3.554\pm  0.101\pm  0.052\\
\delta_\perp - \delta_0 & 0.184\pm  0.070\pm  0.046 &2.958\pm  0.070\pm  0.046\\\hline
\end{array}
\]
\end{minipage}
\end{table}
Under this transformation, $\cos(\delta_\|-\delta_\perp)$ and $\cos(\delta_\perp-\delta_0)$ (eqn.~(\ref{eqn:obs_cos})) change of sign, meaning that the
two set of parameters
$(\delta_\|-\delta_0,\delta_\perp-\delta_0,\ctwob) \leftrightarrow (-(\delta_\|-\delta_0),\pi-(\delta_\perp-\delta_0),-\ctwob)$
are mathematically equivalent~\cite{steph}. But this is considering the $P$--wave $\Kstarz(892)$ only...\\

A $K\pi$ $S$--wave is known to lie in the $K^*(892)$ region~\cite{lass}. The interference with the main $K\pi$ $P$--wave
$K^*(892)$ is the key to break the strong phases ambiguity. Taking into account a $B\to\jpsi(K\pi)_{S\rm{-wave}}$
amplitude, in addition to the three $B\to\jpsi(K\pi)_{P\rm{-wave}}$ ones ($A_0, A_\|, A_\perp$),
introduces the relative strength of the $P$ and $S$ contributions and a new relative phase $\gamma=\delta_S-\delta_0$. There is still an ambiguity:
\begin{equation}
(\delta_\|-\delta_0,\delta_\perp-\delta_0,\gamma) \leftrightarrow (-(\delta_\|-\delta_0),\pi-(\delta_\perp-\delta_0),-\gamma),
\label{eqn:ambigS}
\end{equation}
but the ambiguity on $\gamma$ can be broken.

According the Wigner's causality principle~\cite{wigner}, the phase of a resonance rotates counterclockwise with increasing mass.
In the $K^*(892)$ region, the $(K\pi)_{S\rm{-wave}}$ phase moves slow, while the
$(K\pi)_{P\rm{-wave}}$ phase moves rapidly. The phase $\gamma=\delta_S-\delta_0$ must then rotates clockwise in the  $K^*(892)$ region. Figure~\ref{fig:ambig} shows the $P$ and $S$ wave
intensities as function of the $K\pi$ mass, as well as $\gamma$ where the open points are for strong phases ``Solution I'' and the full points for ``Solution II''. The
physical variation of $\gamma$ is observed for ``Solution II''. As a cross-check of the phase evolution with mass, the $\gamma$ phase evolution is compared in figure~\ref{fig:ambig}
with the evolution observed in the $Kp\to K\pi(n)$ high statistics LASS experiment~\cite{lass}. The agreement is remarkable.\\
\begin{figure}[h]
\begin{minipage}{0.48\linewidth}
\begin{center}
\psfig{figure=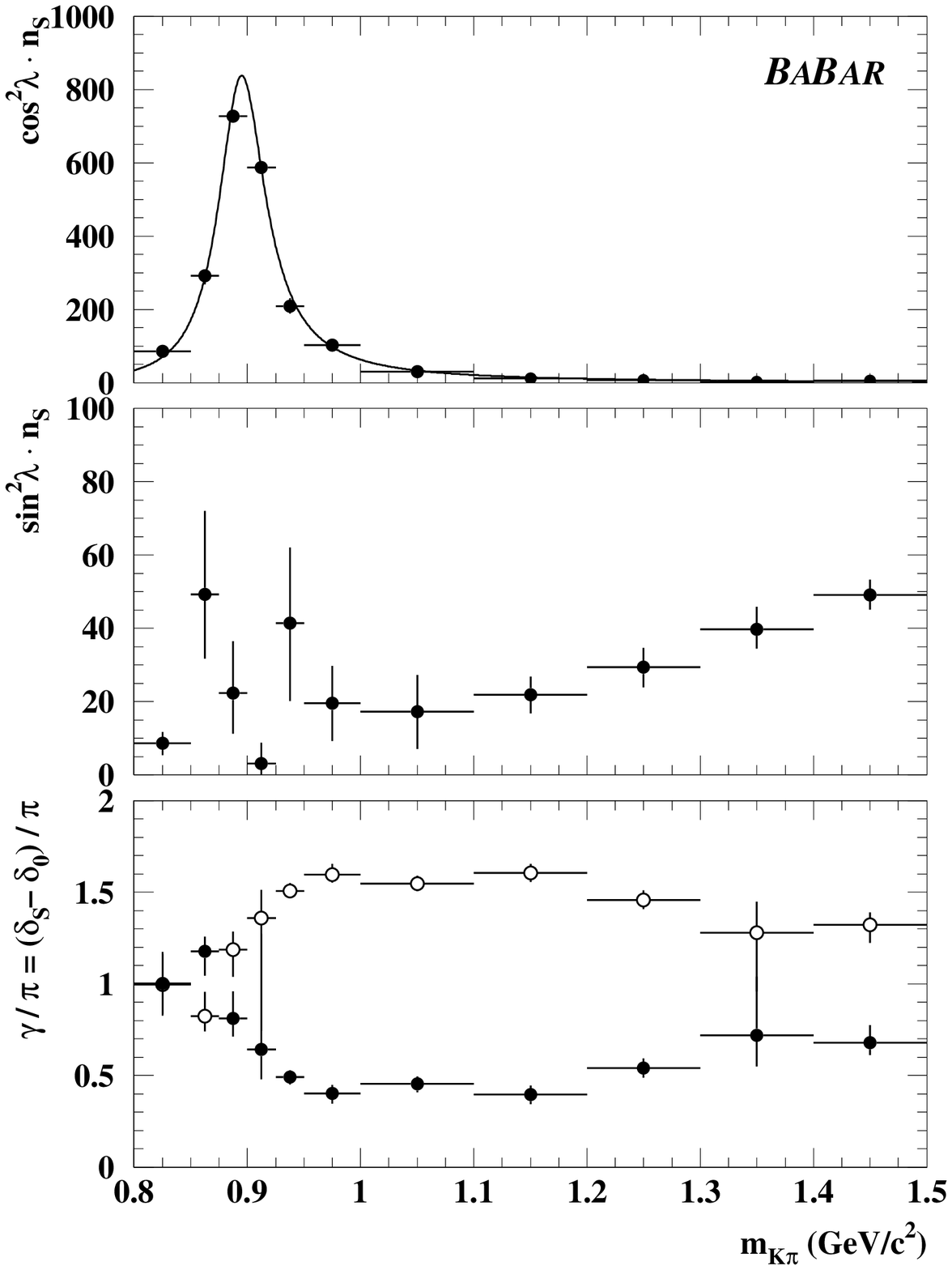,width=0.9\linewidth}
\end{center}
\end{minipage}\hfill
\begin{minipage}{0.48\linewidth}
\begin{center}
\psfig{figure=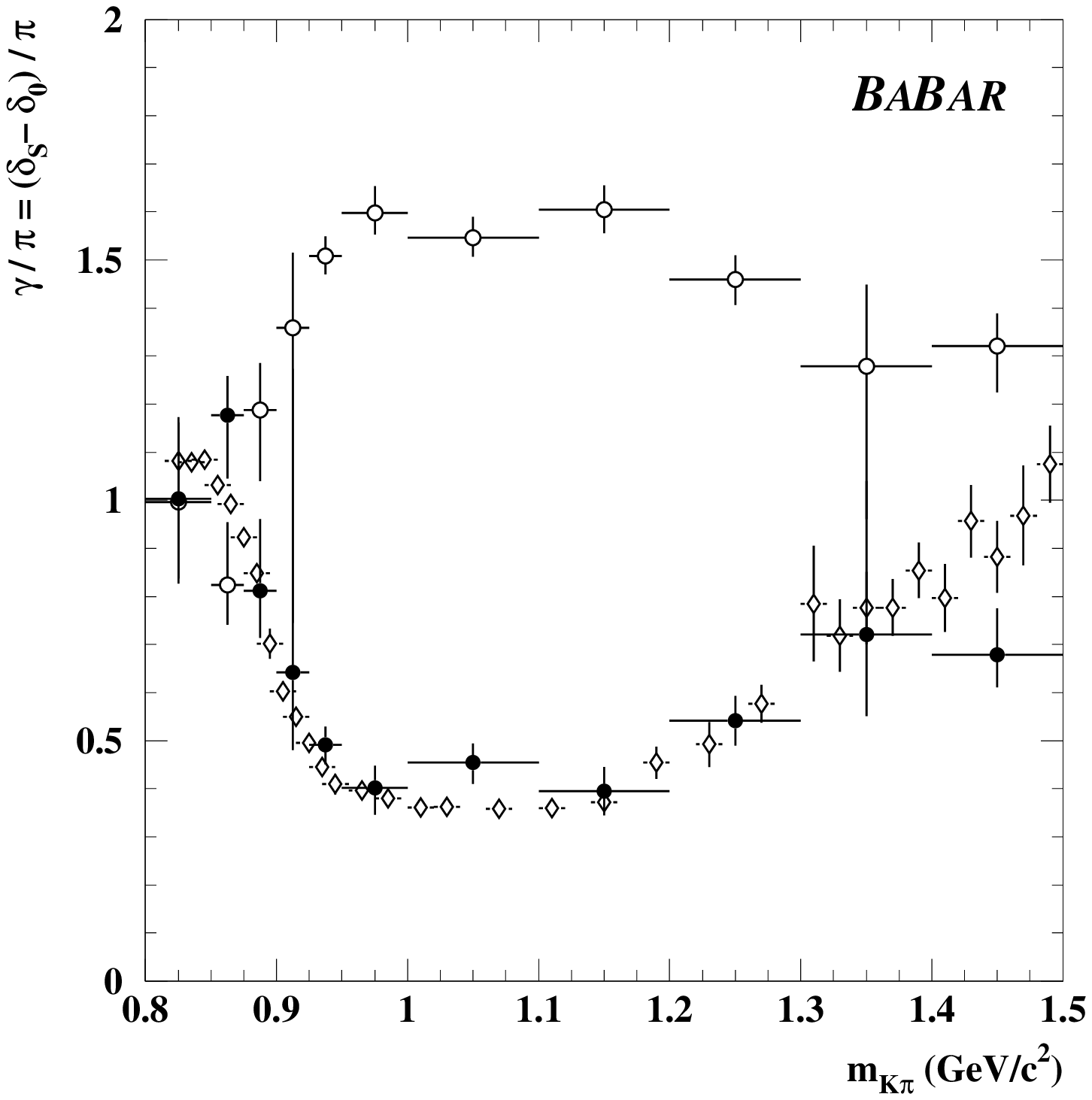,width=0.8\linewidth}
\psfig{figure=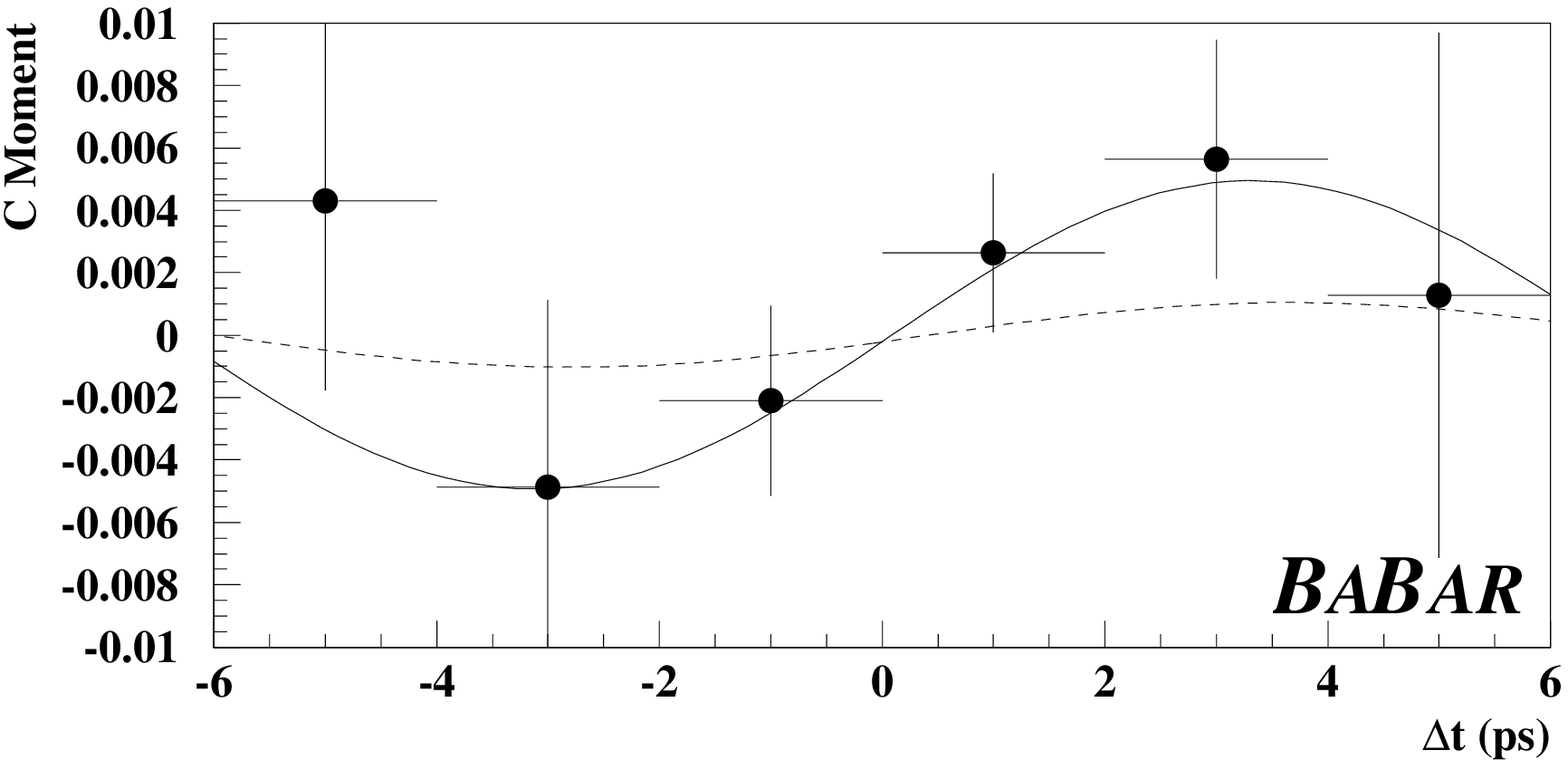,width=0.7\linewidth}
\end{center}
\end{minipage}
\caption{\label{fig:ambig}{\bf Left:} Top(Middle): $P(S)$ wave intensity as function of the $K\pi$ mass ($\Kpm\pimp$ channel only).
The $P$ wave intensity is fitted with a Breit-Wigner lineshape, including centrifugal effect.
Bottom: $\gamma=\delta_S-\delta_0$ as function of the $K\pi$ mass, where the
open points are obtained with strong phases ``Solution I'' (tab.~\ref{tab:amp}) and the full points with ``Solution II'' (tab.~\ref{tab:amp}).
{\bf Right, top:} Comparison of the $\gamma=\delta_S-\delta_0$ phase with the $K\pi$ mass, as in previous plots, with the LASS data, shown as diamond markers.
These data correspond to the isospin $1/2$ contribution, i.e., the one existing
in the $B\to\jpsi(K\pi)$ decay. A global $\pi$ offset was added to the LASS data, which obviously does not change the slope.
{\bf Right, bottom:} Moment of the angular function weighting the \ctwob
contribution in the time and angular dependant distribution. The full line corresponds to $\ctwob=+3.32$, the dashed line to $\ctwob=\sqrt{1-0.731^2}=+0.68$.}
\end{figure}

We perform a time and angular dependant analysis of the $\Bz\to\jpsi(\KS\piz)^{*0}$ sample (104 events), fixing the angular structure of the decay using above amplitude moduli and
strong phases ``Solution II'' (tab.~\ref{tab:amp}). With \stwob and \ctwob free in the fit, we obtain~\cite{steph}
$\stwob=-0.10\pm 0.57({\rm stat})\pm 0.14({\rm syst})$ and 
$\ctwob=+3.32^{+0.76}_{-0.96}({\rm stat})\pm 0.27({\rm syst})$.
Using the world average $\stwob=0.731$ value, we obtain
\begin{equation}
\ctwob=+2.72^{+0.50}_{-0.79}({\rm stat})\pm 0.27({\rm syst}).
\end{equation}
We thus measure a {\em positive} \ctwob value, in agreement with the SM expectation. The fit result for \ctwob can be illustrated making the moment of the angular term weighting \ctwob
in the time and angular dependant distribution, as shown on figure right bottom plot of figure~\ref{fig:ambig}.

Assuming \stwob and \ctwob measure the same angle $2\beta$, we estimate on Monte Carlo that we exclude the negative \ctwob solution at $89\%$ CL. This is preliminary estimate.

\clearpage

\section{\CP measurements with penguin dominated modes}
The penguin dominated modes are considered as ``windows'' to NP. In the SM, contributions beyond the leading penguin
may be uneasy to estimate, depending on the channel. The ``effective \stwob'' measured in these channels may then differ from \stwob, but bounds on these differences are known~\cite{bounds}.
The decay rate $\Bz\to f$ to a \CP-eigenstate $f$, with eigenvalue $\eta_f$, is described by:
\begin{equation}
f_{\stackrel{B^0 tag}{\overline{B}^0 tag}}(\Delta t) = \frac{e^{-|\Delta t|/\tau_{B^0}}}{4\tau_{B^0}}\times\left[1\mp\left(C\cos(\Delta m_{B^0}\Delta t) - 
S\sin(\Delta m_{B^0}\Delta t)\right)\right]
\label{eqn:sc}
\end{equation}
where $\Delta t$ is the time difference between the decays of  the $B$ meson studied and the other $B$ meson ($B_{tag}$), which decay products are used
in a partial reconstruction to infer its $\Bz$ or $\Bzb$
flavor. In a simplistic case, $C=0$ and $S=-\eta_f\stwob$.

\subsection{\Bz\to$\phi$\KS and \Bz\to$\phi$\KL}
The decay \Bz\to$\phi$\Kz is a $b\to s\overline{s}s$ quark level decay. In the SM, the expected asymmetry
$S_{\phi\KS}(S_{\phi\KL})$ is very close to $\sim +\stwob(-\stwob)$. The \babar\ 
measurement is reported in table~\ref{tab:sc} and decay rates shown in figure~\ref{fig:phiK-mes-de}.  The measurement is
in agreement with the SM expectation.
\begin{figure}[h]
\begin{minipage}{0.35\linewidth}
\begin{center}
\psfig{figure=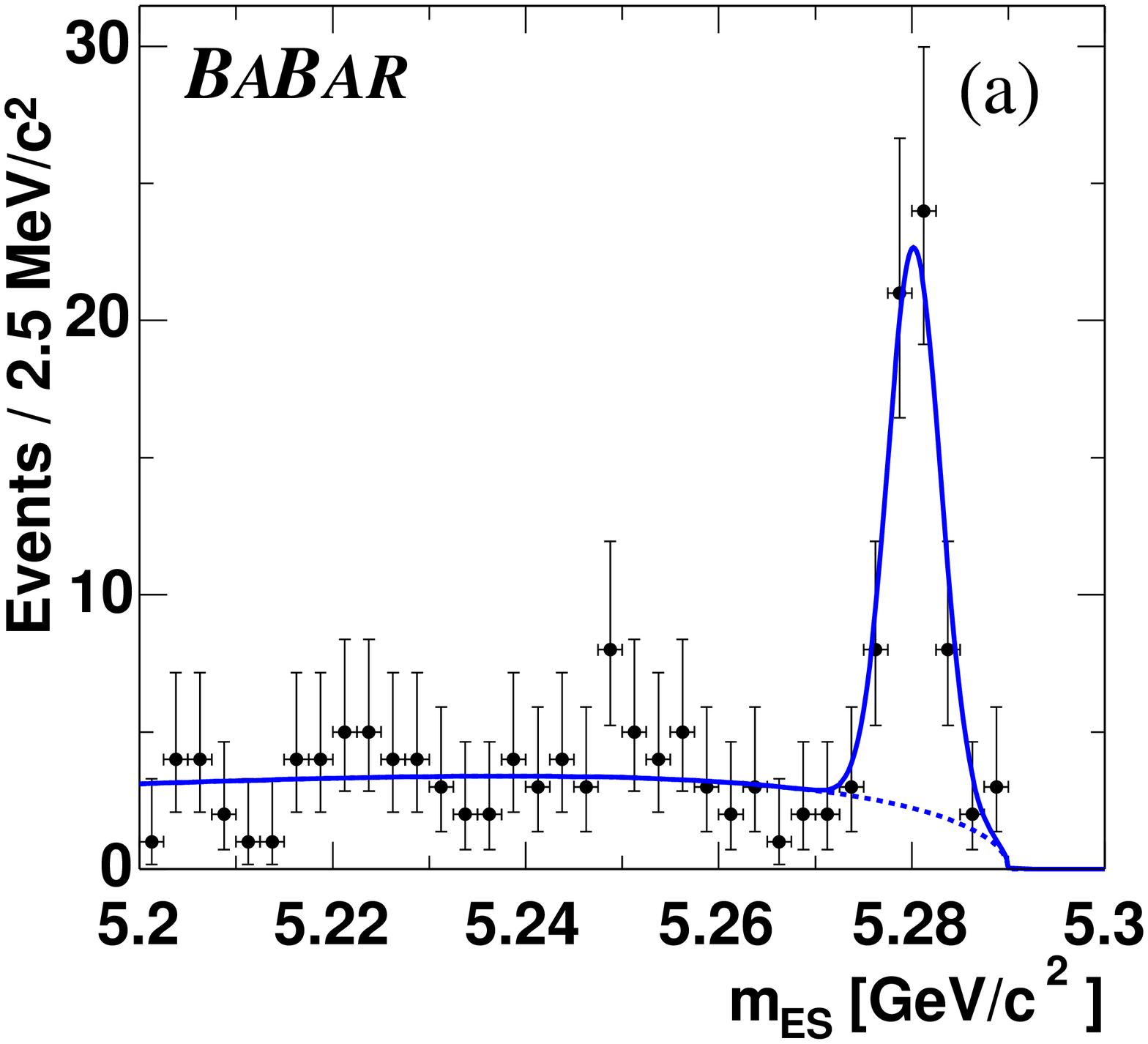,width=0.8\linewidth}
\psfig{figure=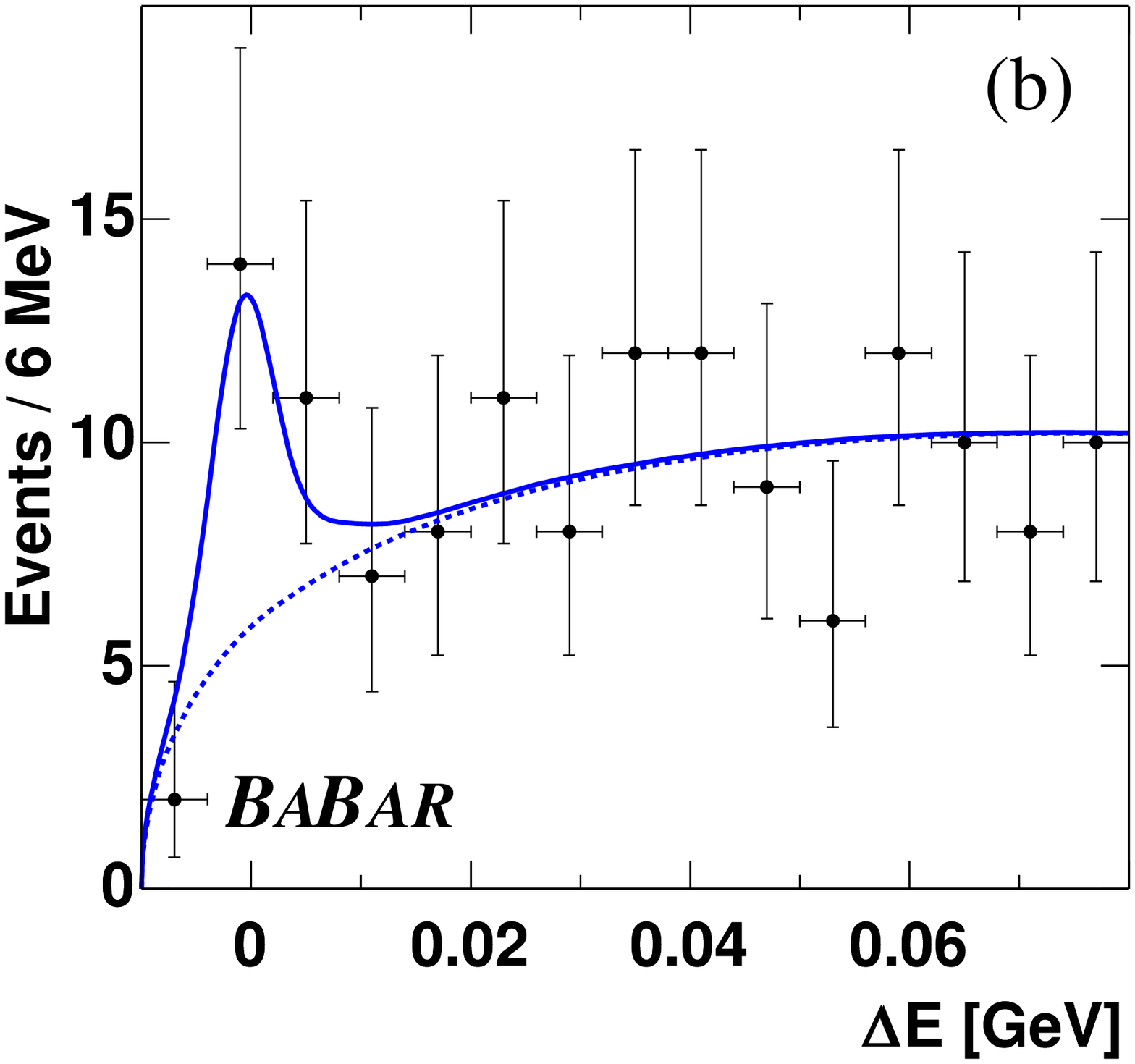,width=0.8\linewidth}
\end{center}
\end{minipage}\hfill
\begin{minipage}{0.50\linewidth}
\begin{center}
\psfig{figure=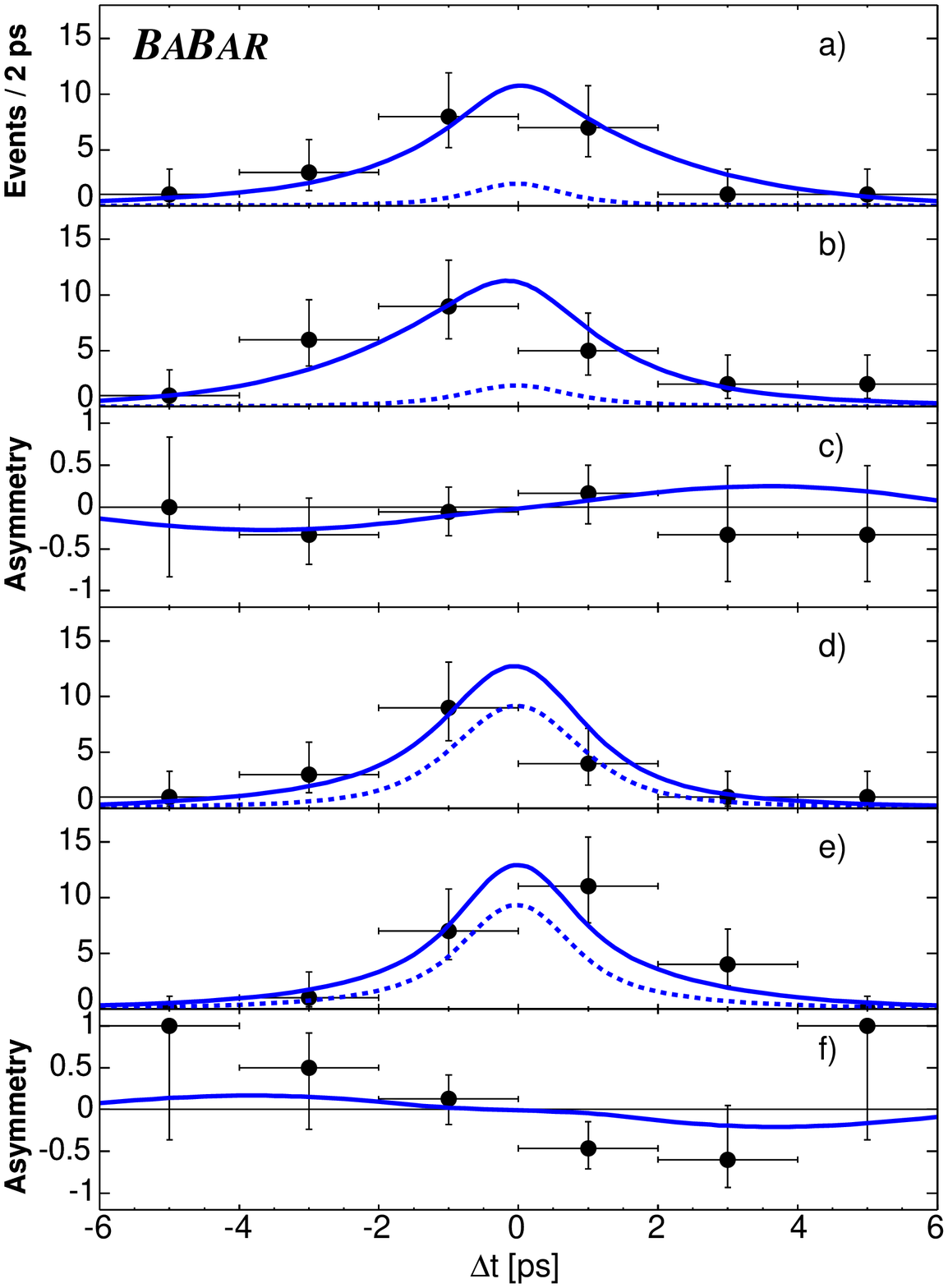,width=0.8\linewidth}
\end{center}
\end{minipage}
\caption{\label{fig:phiK-mes-de}{\bf Left:} Top: Mass distribution of $\Bz\to\phi\KS$ events, with $70\pm 9$ candidates found out of a $108$fb$^{-1}$ sample. Bottom:
Difference of measured and expected energy of $\Bz\to\phi\KL$ candidates: only the \KL direction is known and the $B$ mass constraint is used to estimate the $B$ energy.
$52\pm 16$ candidates are found.
{\bf Right:} $\Delta t$ distributions and asymmetry for $\Bz\to\phi\KS$ (three top plots) and $\Bz\to\phi\KL$ (three bottom plots).}
\end{figure}

\subsection{\Bz\to\Kp\Km\KS and \Bpm\to\Kpm\KS\KS}
The \CP asymmetry of the quark level decay  $b\to s\overline{s}s$ can also be measured with the inclusive \Bz\to\Kp\Km\KS decay (excluding $\phi$\to\Kp\Km)
and benefits from larger statistics than the \Bz\to$\phi$\KS mode (see fig.~\ref{fig:KKK}).
In contrast with  \Bz\to$\phi$\KS, the \CP content is not known {\it a priori}. It can be determined from $B\to KKK$
branching ratios of charged and neutral $B$ mesons~\cite{belle} as:
$f_{even}={2\Gamma(\Bp\to\Kp\KS\KS)}/{\Gamma(\Bz\to\Kp\Km\KS)}$.
\babar\ measures the following branching ratios $Br(\Bp\to\Kp\KS\KS) = (10.7\pm1.2\pm1.0)\times 10^{-6}$ and $Br(\Bz\to\Kp\Km\KS )=(23.8\pm 2.0\pm 1.6)\times 10^{-6}$, and obtains
$f_{even}= 0.98\pm 0.15\pm 0.04$,
which is compatible with a pure \CP even state. In the SM the expected \Bz\to\Kp\Km\KS \CP asymmetry if then $S_{\Kp\Km\KS}\sim-\stwob$.
The \CP asymmetry parameters $S$ and $C$ measured are shown in table~\ref{tab:sc} and the decay rates are shown in figure~\ref{fig:KKK}.
The first measurement of the \CP-violating charge asymmetry, $A_{\CP}$, in the \Bpm\to\Kpm\KS\KS decay is also made:
$A_{\CP}(\Bpm\to\Kpm\KS\KS)=-0.042\pm 0.114({\rm stat})\pm 0.02({\rm syst})$.

\begin{figure}[h]
\begin{minipage}{0.3\linewidth}
\psfig{figure=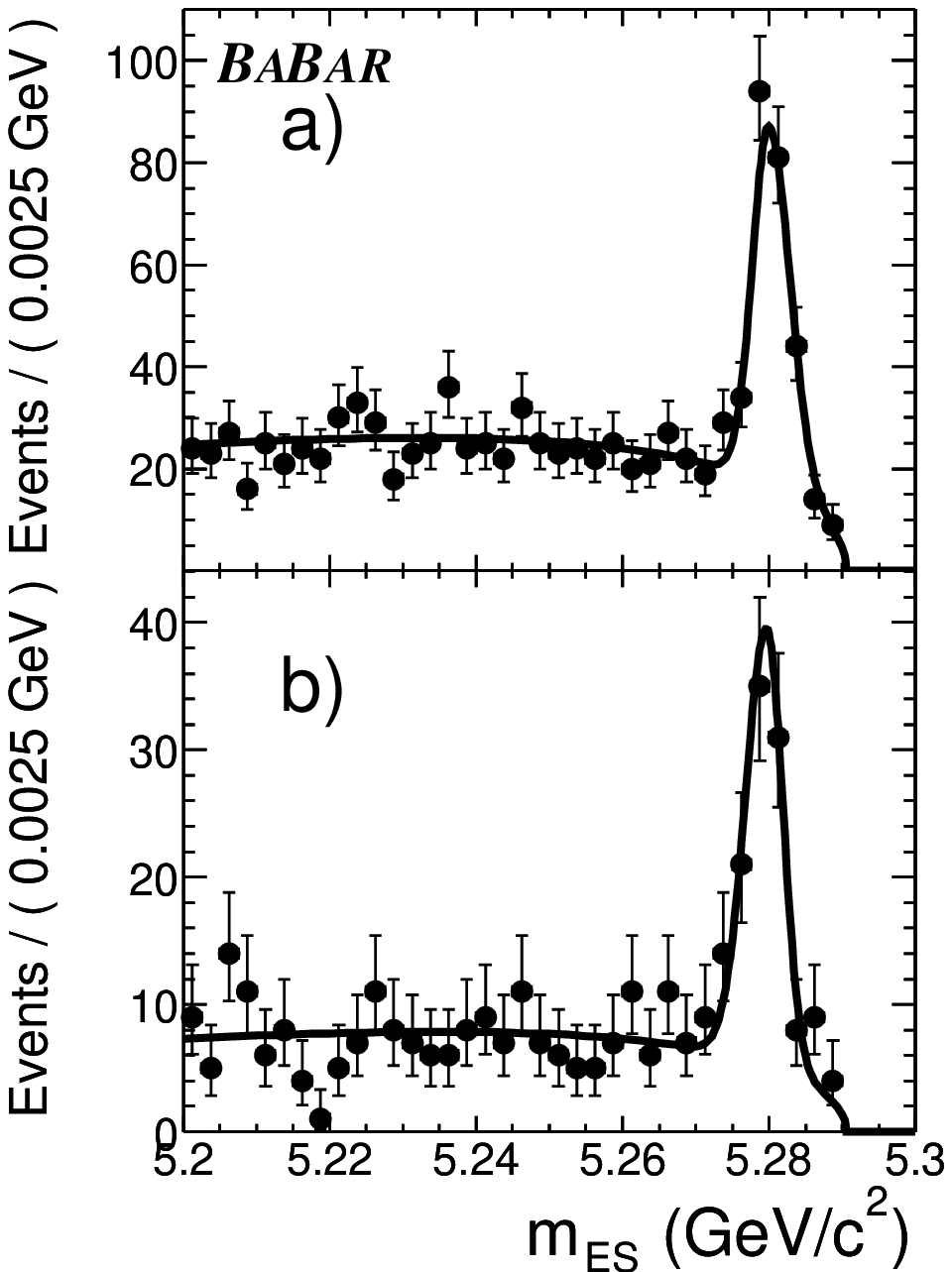,width=\linewidth}
\end{minipage}\hfill
\begin{minipage}{0.48\linewidth}
\psfig{figure=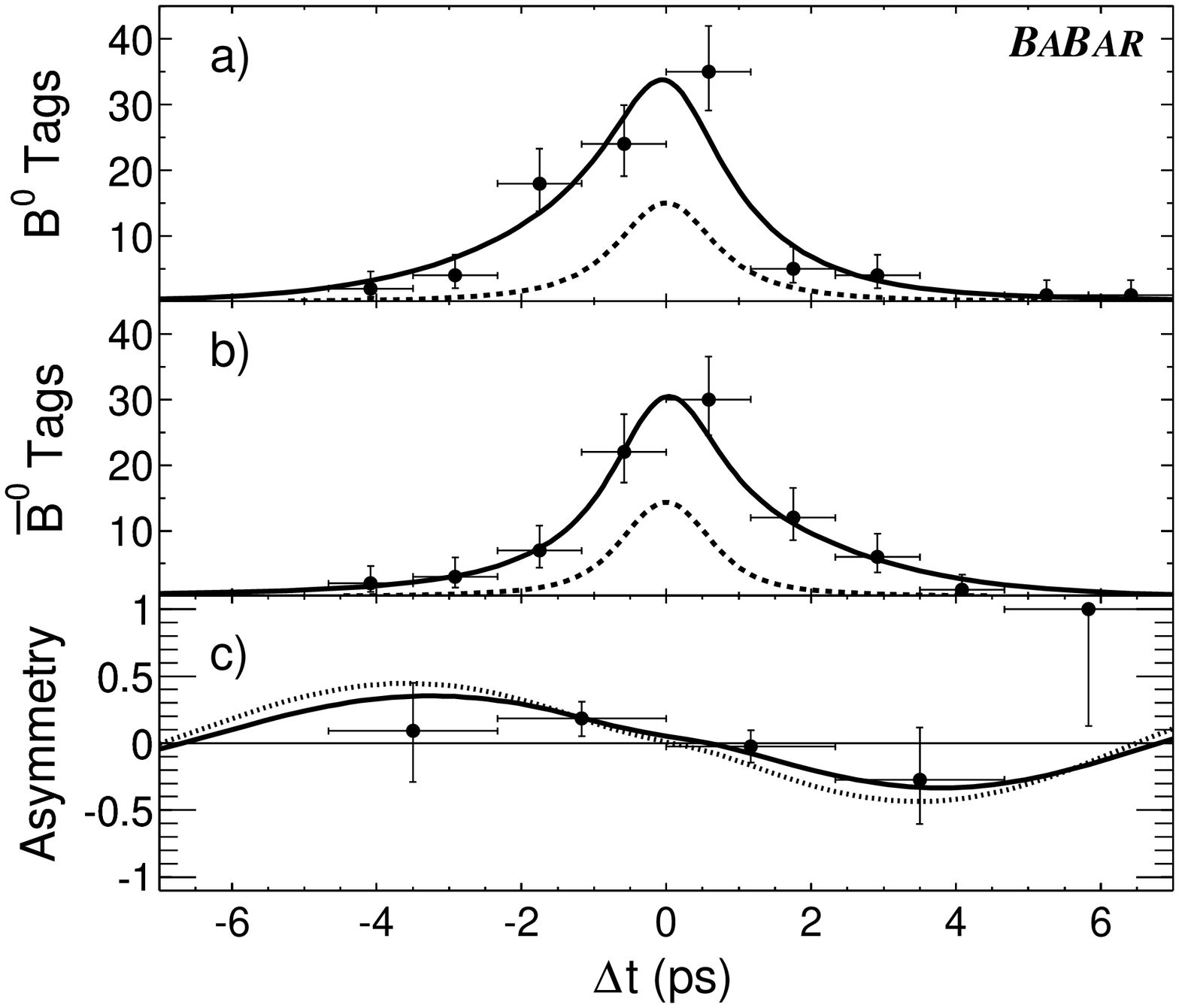,width=\linewidth}
\end{minipage}
\caption{\label{fig:KKK}{\bf Left:} Mass distribution of a) \Bz\to\Kp\Km\KS and b) \Bpm\to\Kpm\KS\KS candidates. $201\pm 16$ and $122\pm 14$ candidates are found
respectively out of
a 111fb$^{-1}$ sample. {\bf Right:} $\Delta t$ distributions and asymmetry.}
\end{figure}

\subsection{\Bz\to\piz\KS}
This a $b\to s\overline{d}d$ quark level decay. The SM expectation for $S_{\piz\KS}$ is $\sim+\stwob$. An experimental issue with this decay is the determination
of the $\Bz$ vertex: no charged particules emerge from the $\Bz$ vertex. The reconstructed $\KS$ direction, with a beam spot constraint in the plane transverse to
the beam direction, is used to estimate the vertex position.
This vertex determination technique is checked with $\Bz\to\jpsi\KS$ and $\Bp\to\pip\KS$ decays, ignoring the $\jpsi$ or the $\pip$. It is
also checked by measuring the $\Bz$ lifetime. The \CP asymmetry measurements are shown in table~\ref{tab:sc} and decay rates plots are shown in figure~\ref{fig:piK-deltat}.

\begin{figure}[h]
\begin{minipage}{0.42\linewidth}
\begin{center}
\psfig{figure=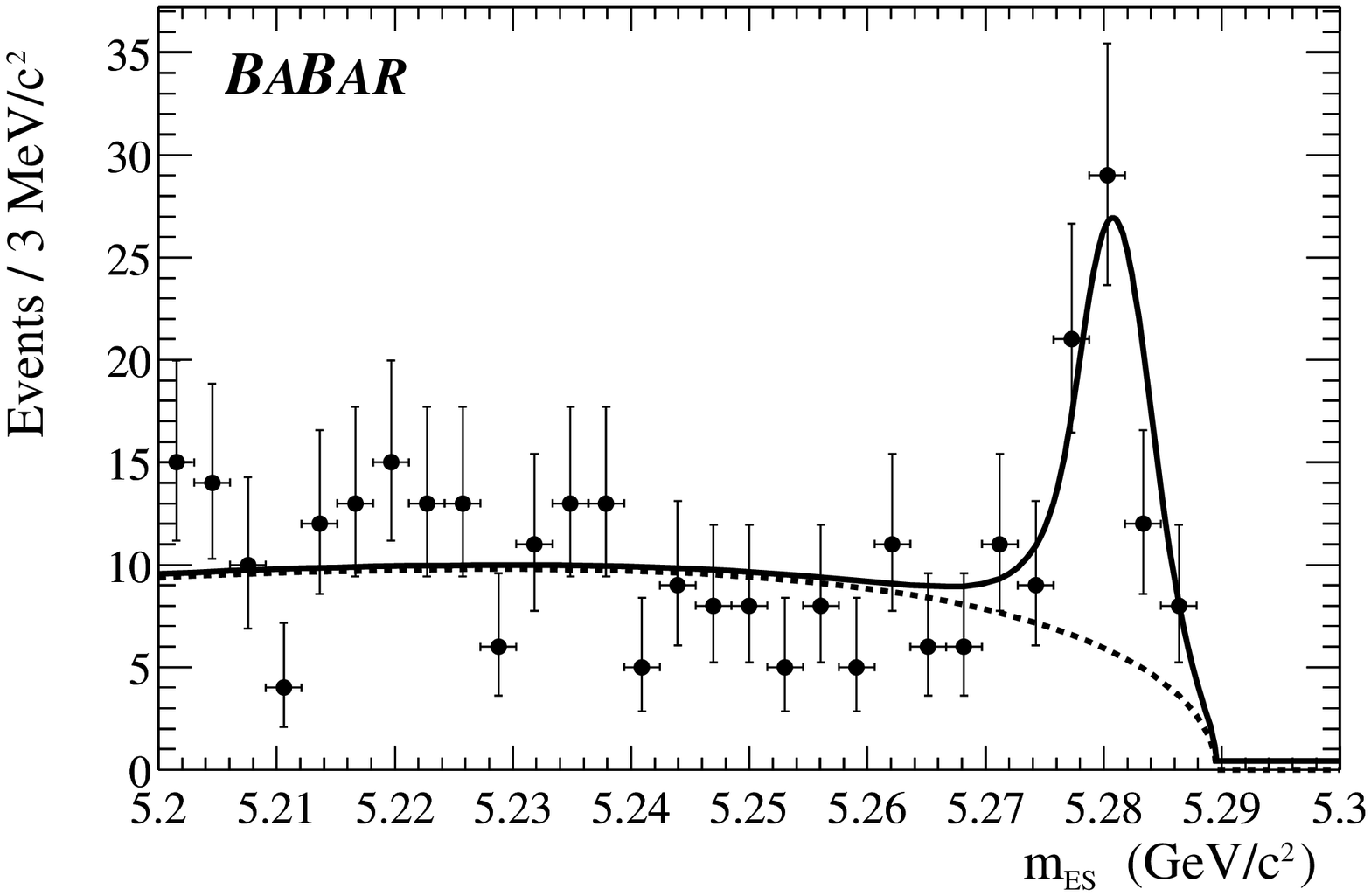,width=\linewidth}
\end{center}
\end{minipage}\hfill
\begin{minipage}{0.42\linewidth}
\begin{center}
\psfig{figure=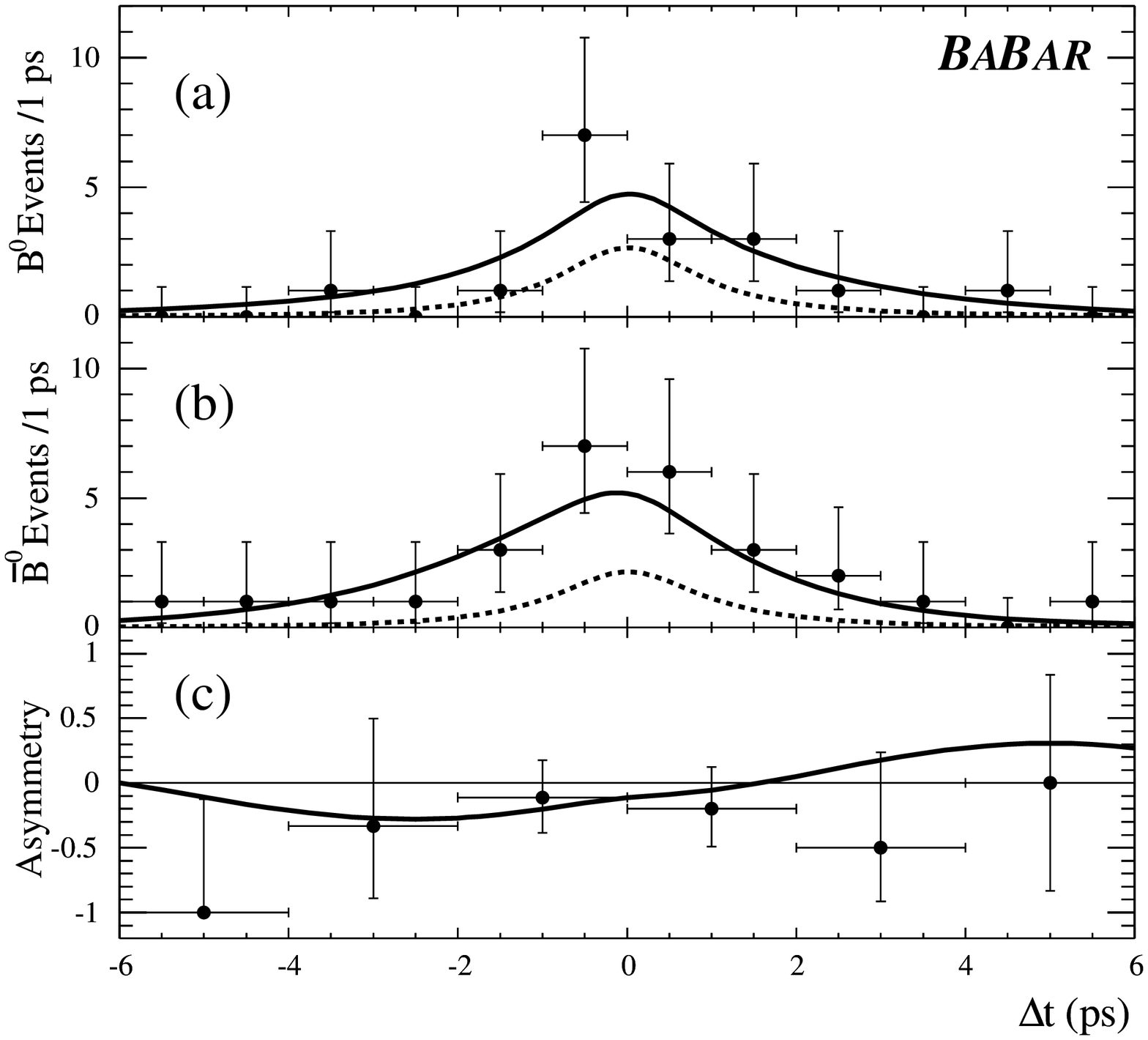,width=\linewidth}
\end{center}
\end{minipage}
\caption{\label{fig:piK-deltat}{\bf Left:} Mass distribution of \Bz\to\piz\KS candidates, with $122\pm 16$ candidates found out of a 113fb$^{-1}$ sample.
{\bf Right:} $\Delta t$ distributions and asymmetry.}
\end{figure}

\subsection{\Bz\to\fz$(980)$\KS, \fz$(980)$\to\pip\pim}
The structure of the scalar meson \fz$(980)$ is obscure, but recent studies favor an usual $q\overline{q}$ interpretation~\cite{fzero}, with
$f_0 = \cos\phi_s s\overline{s} + \sin\phi_s\left(u\overline{u}+d\overline{d}\right)/\sqrt{2}$ and $\phi_s=(-48\pm 6)^{\rm o}$.
The decay \Bz\to\fz$(980)$\KS should then be dominated by the $b\to s\overline{s}s$ penguin, since the $s\overline{s}$ component is sizeable and the
$b\to u\overline{u}s$ tree is doubly Cabbibo suppressed compared to the leading penguin. The \Bz\to\fz$(980)$\KS \CP asymmetry expected in the SM
is then $\sim -\stwob$.

A quasi two-body analysis is performed, with a cut in the $\pi\pi K$ Dalitz plot made to reduce the contributions from the $\rho_0$ and the $f_0(1370)$.
This is the first observation of the \Bz\to\fz$(980)$\KS decay (fig.~\ref{fig:f0K}). The signal is checked verifying that a fit to
 the $\pip\pim$ mass spectrum (fig.~\ref{fig:f0K}) with a relativistic Breit-Wigner leads to a mass and a width compatible with the $f_0$ PDG values.
The \CP fit result is shown in table~\ref{tab:sc}, with decay rates distributions shown in figure~\ref{fig:f0K}.

\begin{figure}[h]
\begin{minipage}{0.35\linewidth}
\begin{center}
\psfig{figure=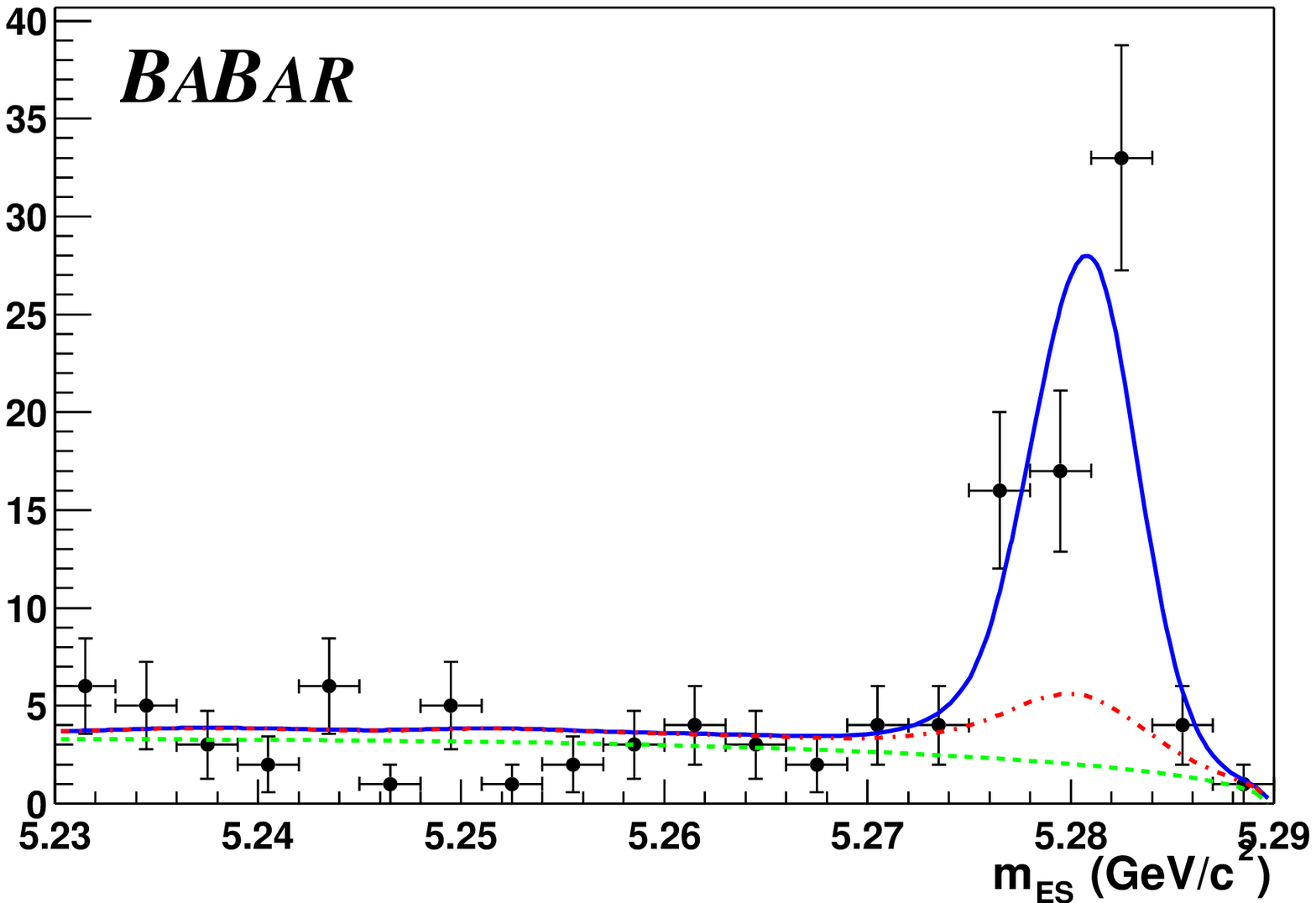,width=\linewidth}
\psfig{figure=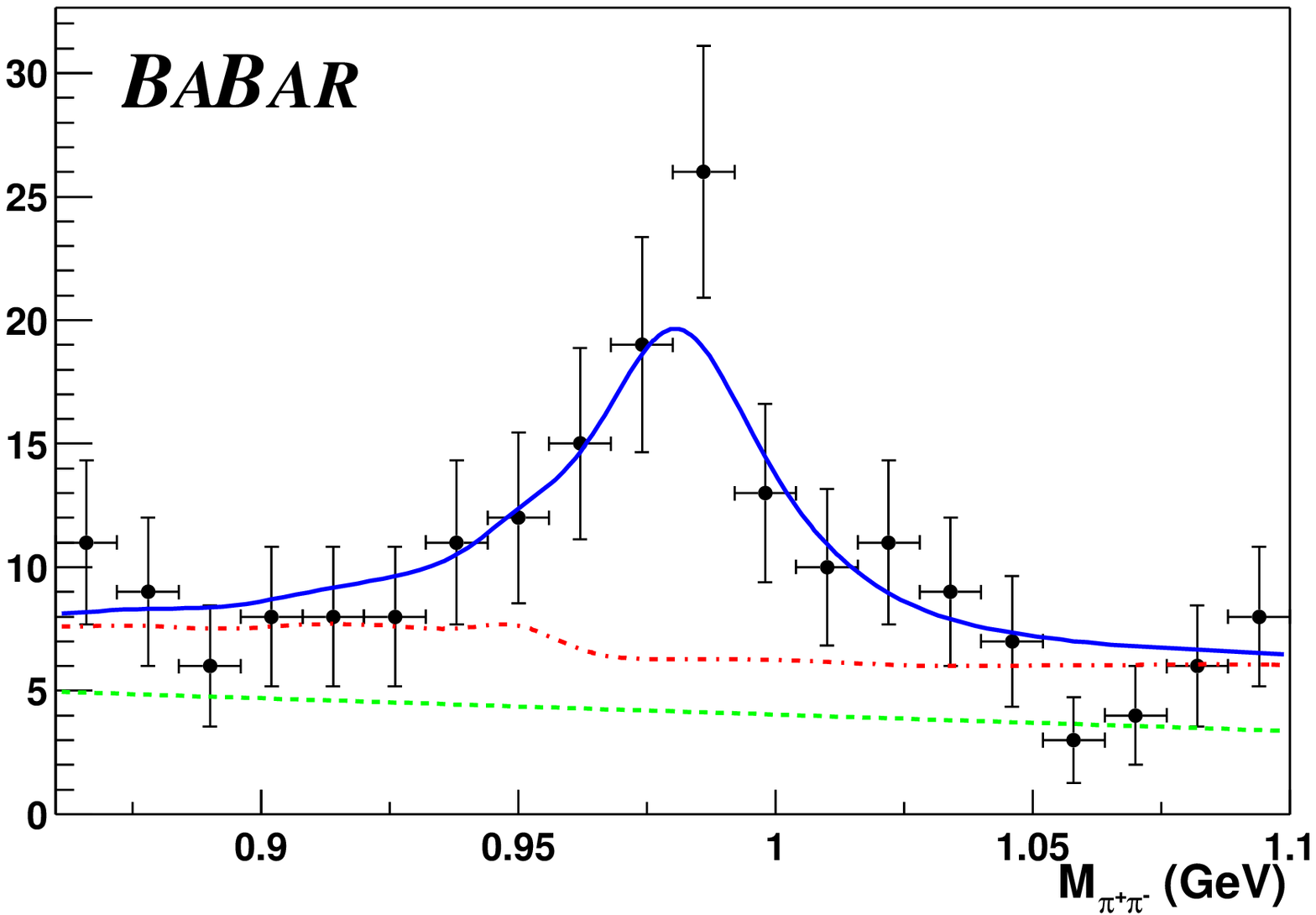,width=\linewidth}
\end{center}
\end{minipage}\hfill
\begin{minipage}{0.50\linewidth}
\begin{center}
\psfig{figure=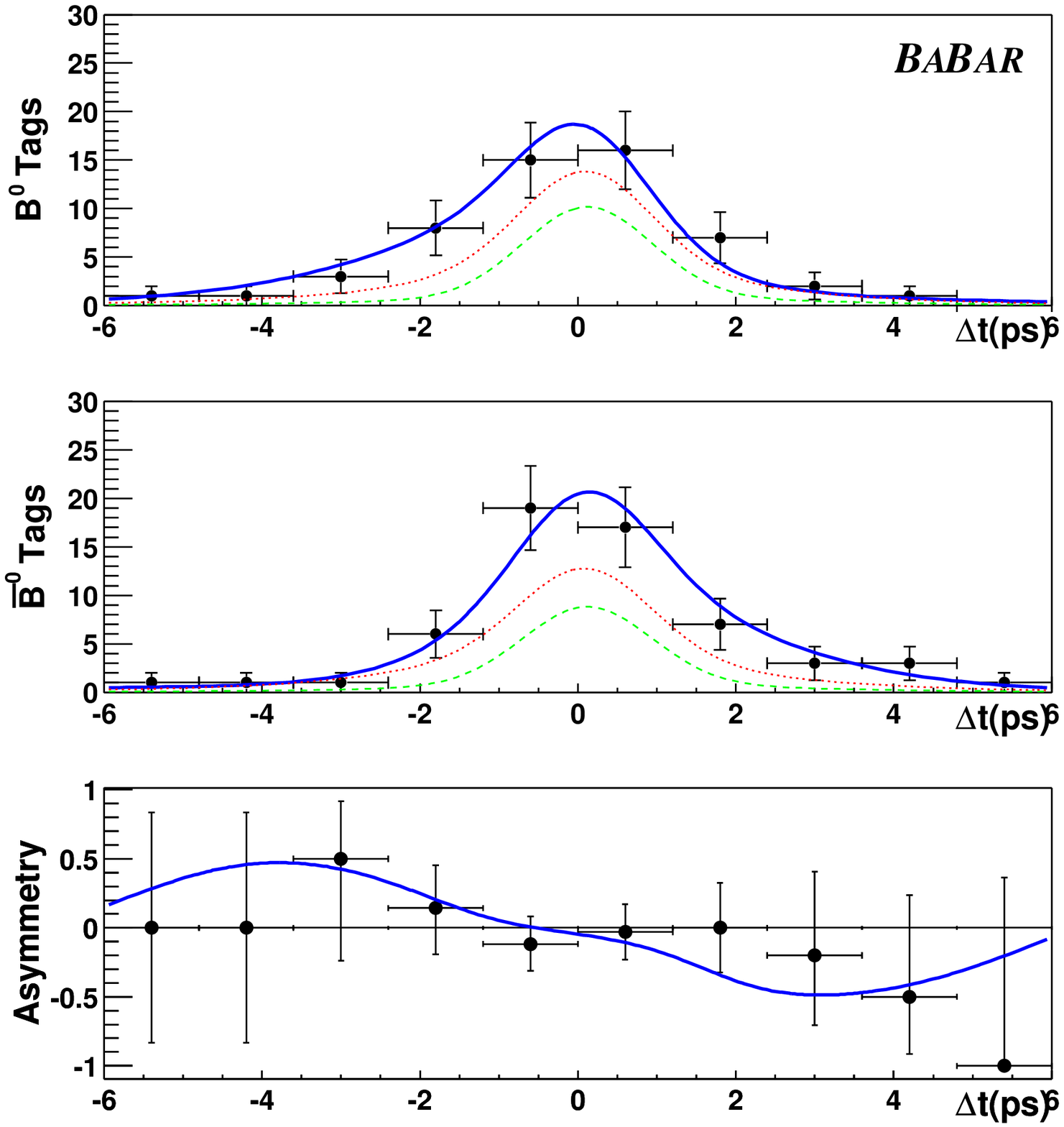,width=\linewidth}
\end{center}
\end{minipage}
\caption{\label{fig:f0K}{\bf Left:} Top: Mass distribution of \Bz\to\fz$(980)$\KS candidates. $94\pm 14 \pm 6$ out of a 111fb$^{-1}$ sample are found.
Bottom: $\pip\pim$ invariant mass, fitted with a relativistic Breit-Wigner function. {\bf Right:} $\Delta t$ distributions and asymmetry.}
\end{figure}

\begin{table}
\begin{center}
\caption{\label{tab:sc}$S$ and $C$ \CP parameters (eqn.~\ref{eqn:sc}) measured for various $B$ decay modes. The (approximative) SM expectation for $S$ is given in the
second column. For $S$ and $C$, the first uncertainty is statistical,
the second one systematical. The ``$f_{even}$'' uncertainty for $S$ of  $\Kp\Km\KS$ comes from the uncertainty on $f_{even}$ itself.
Details for each mode are given in the text.}
\begin{tabular}{|l|c|c|c|}\hline
$B$ decay         & SM exp. & $S$ & $C$ \\\hline
  $\phi\Kz$       & $+\stwob$ & $+0.47\pm0.34^{+0.08}_{-0.06}$                 & $+0.10\pm0.33\pm0.10$          \\
  $\Kp\Km\KS$     & $-\stwob$ & $-0.56\pm 0.25\pm 0.04^{+0}_{-0.17}(f_{even})$ & $-0.10\pm 0.19\pm 0.09$        \\
  $\piz\KS$       & $+\stwob$ & $+0.48^{+0.38}_{-0.47}\pm 0.11$                & $+0.40^{+0.27}_{-0.28}\pm 0.10$ \\
  $\fz(980)\KS$   & $-\stwob$ & $-1.62^{+0.56}_{-0.51}\pm 0.10$                & $+0.27\pm 0.36\pm 0.12$\\\hline
\end{tabular}
\end{center}
\end{table}

\section{Conclusion}
A novel method to resolve the ambiguity of the strong phases in the $B\to\jpsi K^*$ has been used.
It allows to measure the sign of \ctwob with \Bz\to\jpsi\Kstar;\Kstarz\to\KS\piz
free from the strong phases ambiguity. This sign is found positive, in agreement with SM expectation. Measurements of \CP asymmetries in the penguin dominated modes
\Bz\to$\phi$\Kz, \Bz\to\Kp\Km\KS, \Bz\to\piz\KS and \Bz\to\fz$(980)$\KS are all found compatible with SM expectations. At the present level of statistics,
\babar's picture of $\beta$ is SM-like.

\end{document}